\documentclass[a4paper,twocolumn,showpacs,superscriptaddress]{revtex4}

\usepackage[english]{babel}
\usepackage{graphicx}
\usepackage{xspace}
\usepackage{longtable}
\usepackage{amsmath,amssymb}

\DeclareMathAlphabet{\mathbfit}{OT1}{cmr}{bx}{it}

\begin{document}

\title{Sr$_2$CrOsO$_6$: Endpoint of a spin polarized metal-insulator
transition by 5$d$ band filling}

\author{Y.~Krockenberger}
\affiliation{Darmstadt University of Technology, Petersenstr.~23,
64287 Darmstadt, Germany} \affiliation{Max-Planck-Institute for
Solid State Research, Heisenbergstr.~1, 70569 Stuttgart, Germany}
\author{K.~Mogare}
\affiliation{Max-Planck-Institute for Solid State Research,
Heisenbergstr.~1, 70569 Stuttgart, Germany}
\author{M.~Reehuis}
\affiliation{Max-Planck-Institute for Solid State Research,
Heisenbergstr.~1, 70569 Stuttgart, Germany}
\affiliation{Hahn-Meitner-Institut (HMI), 14109 Berlin, Germany}
\author{M.~Tovar}
\affiliation{Hahn-Meitner-Institut (HMI), 14109 Berlin, Germany}
\author{M.~Jansen}
\affiliation{Max-Planck-Institute for Solid State Research,
Heisenbergstr.~1, 70569 Stuttgart, Germany}
\author{G.~Vaitheeswaran}
\affiliation{Max-Planck-Institute for Solid State Research,
Heisenbergstr.~1, 70569 Stuttgart, Germany}
\affiliation{Department of Materials Science and Engineering,
Royal Institute of Technology (KTH), 10044 Stockholm, Sweden}
\author{V. Kanchana}
\affiliation{Max-Planck-Institute for Solid State Research,
Heisenbergstr.~1, 70569 Stuttgart, Germany}
\affiliation{Department of Materials Science and Engineering,
Royal Institute of Technology (KTH), 10044 Stockholm, Sweden}
\author{F. Bultmark}
\affiliation{Department of Physics, University of Uppsala
Box 530, 75121 Uppsala, Sweden}
\author{A. Delin}
\affiliation{Department of Materials Science and Engineering, Royal
Institute of Technology (KTH), 10044 Stockholm, Sweden}
\author{F.~Wilhelm}
\affiliation{European Synchrotron Radiation Facility (ESRF), 6 Rue
Jules Horowitz, BP 220, 38043 Grenoble Cedex 9, France}
\author{A.~Rogalev}
\affiliation{European Synchrotron Radiation Facility (ESRF), 6 Rue
Jules Horowitz, BP 220, 38043 Grenoble Cedex 9, France}
\author{A.~Winkler}
\affiliation{Darmstadt University of Technology, Petersenstr.~23,
64287 Darmstadt, Germany}
\author{L.~Alff}
\email{alff@oxide.tu-darmstadt.de} \affiliation{Darmstadt University
of Technology, Petersenstr.~23, 64287 Darmstadt, Germany}

\date{Received 13 December 2006}

\begin{abstract}
In the search for new spintronic materials with high
spin-polarization at room-temperature, we have synthesized an
osmium based double perovskite with a Curie-temperature of 725\,K.
Our combined experimental results confirm the existence of a
sizable induced magnetic moment at the Os site, supported by
band-structure calculations in agreement with a proposed kinetic
energy driven mechanism of ferrimagnetism in these compounds. The
intriguing property of Sr$_2$CrOsO$_6$ is that it is at the
endpoint of a metal-insulator transition due to 5$d$ band filling,
and at the same time ferrimagnetism and high-spin polarization is
preserved.
\end{abstract}
\maketitle

A so-called half-metal is a highly desired material for
spintronics, as only charge carriers having one of the two
possible polarization states contribute to conduction. In the
class of the ferrimagnetic double perovskites such half-metals are
well known, e.~g.~Sr$_2$FeMoO$_6$ \cite{Kobayashi:98}. The here
described compound Sr$_2$CrOsO$_6$ is special, as it has a
completely filled $5d$ $t_{\text 2g}$ minority spin orbital, while
the majority spin channel is still gapped. It is thus at the
endpoint of an ideally fully spin-polarized metal-insulator
transition. At the metallic side of this transition we have the
half-metallic materials Sr$_2$CrWO$_6$ \cite{Philipp:03} and
Sr$_2$CrReO$_6$ \cite{Kato:02,DeTeresa:05}. Within the unique
materials class of double perovskites, therefore, one can find
high Curie-temperature ferrimagnets with spin-polarized
conductivity ranging over several orders of magnitude from
ferrimagnetic metallic to ferrimagnetic insulating {\em tunable}
by electron doping. Note that Sr$_2$CrOsO$_6$, where a regular
spin polarized 5$d$ band is shifted below the Fermi level, is
fundamentally different from a diluted magnetic semiconductor,
where spin-polarized charge carriers derive from impurity states.

While for simple perovskites as the half-metallic ferromagnetic
manganites the Curie-temperature, $T_\textrm{C}$, is in the
highest case still close to room-temperature, half-metallic
ferrimagnetic double perovskites can have a considerably higher
$T_\textrm{C}$ \cite{Serrate:07}. It has been suggested that
ferrimagnetism in the double perovskites is kinetic energy driven
\cite{Sarma:00,Kanamori:01,Fang:01}. In short, due to the
hybridization of the exchange split $3d$-orbitals of Fe$^{3+}$
($3d^{\,5}$, spin majority orbitals fully occupied) or Cr$^{3+}$
($3d^{\,3}$, only $t_{\text{2g}}$ are fully occupied), and the
non-magnetic $4d$/$5d$-orbitals of Mo, W, Re or Os ($N$-sites), a
kinetic energy gain is only possible for the minority spin
carriers. This will lead to a corresponding shift of the bare
energy levels at the non-magnetic site, and a strong tendency to
half-metallic behavior. This mechanism is operative for the
Fe$^{3+}$ and Cr$^{3+}$ ($M$ sites) compounds \cite{Philipp:03},
where all $3d$ majority spin states resp.~all $t_{\text{2g}}$
majority spin sates are fully occupied and represent localized
spins. In agreement with band-structure calculations
\cite{Kobayashi:98,Sarma:00,Jeng:03,Philipp:03,Vaitheeswaran:05,Vaitheeswaran:06}
this mechanism is naturally associated with half-metallic
behavior, as the spin-polarized conduction electrons mediate
antiferromagnetic order between $M$ and $N$ ions, and thus,
ferromagnetic order between the $M$ sites. In addition, this
mechanism will lead to an {\em induced} magnetic moment at the
non-magnetic sites as Mo, W, Re, or Os, in convincing quantitative
agreement with recent band-structure calculations. It has been
suggested and verified by observation of the corresponding x-ray
magnetic circular dichroism (XMCD) that the induced magnetic
moment at the non-magnetic site scales with $T_\textrm{C}$ of the
compound \cite{Majewski:05apl,Majewski:05prb,Sikora:06}, or -- in
other words -- the band-width in the conducting spin minority
channel scales with the magnetic transition temperature. This
behavior has also been confirmed qualitatively by nuclear magnetic
resonance (NMR) experiments for FeRe-based double perovskites
\cite{Wojcik:05}. Quantitatively, however, Re shows an unusually
enhanced induced magnetic moment in the double perovskite
structure, as compared to Mo, W, and Os.

An obvious way to change the band-filling is by doping at the
alkaline earth site. Substituting for example Sr$^{2+}$ by
La$^{3+}$ results in electron doping. For the ferrimagnetic double
perovskites it has been shown that increasing the band-width by La
doping generically increases $T_\textrm{C}$
\cite{Navarro:01,Frontera:03,Gepraegs:06}. The increase is
substantial, i.e.~more than 100\,K per one doped electron per unit
cell. However, there is an alternative way of doping by using the
$N$ site itself \cite{Alff:06}. Within this line of thinking there
is a natural explanation why Sr$_2$CrReO$_6$ was up to now the
ferrimagnetic double perovskite with the highest
Curie-temperature, 635\,K \cite{Kato:02}. Compared to
Sr$_2$CrWO$_6$ ($T_\textrm{C}\approx 450-500$\,K
\cite{Philipp:01,Philipp:03,Gepraegs:06}) with W$^{5+}$ ($5d^1$),
Sr$_2$CrReO$_6$ with Re$^{5+}$ ($5d^2$) possesses just one
electron more in the spin minority channel. Therefore, it
corresponds to the compound LaSrCrWO$_6$. The increase of
$T_\textrm{C}$ by more than 100\,K in both cases (as estimated
from the case of LaSrFeMoO$_6$ \cite{Navarro:01} and LaCaCrWO$_6$
\cite{Gepraegs:06}) underlines the justification of this
comparison. Following this argument, Sr$_2$CrOsO$_6$ with
Os$^{5+}$ ($5d^3$) becomes a natural candidate for an ordered
ferrimagnetic double perovskite with even higher $T_\textrm{C}$.
After an early attempt to synthesize this compound, it has been
(erroneously) concluded that it is {\em not} ferromagnetic at
room-temperature \cite{Sleight:61}. The CrTa, CrW, CrRe, CrOs
series is an ideal system to study the effect of 5$d$ band
filling.
In order to investigate these ideas we have synthesized phase pure
polycrystalline Sr$_2$CrOsO$_6$ samples and measured the net
magnetization by SQUID magnetometry. The (magnetic) crystal
structure has been extensively studied by neutron scattering and
XMCD.

\begin{figure}[t]
\centering{%
\includegraphics[width=0.9\columnwidth,clip=]{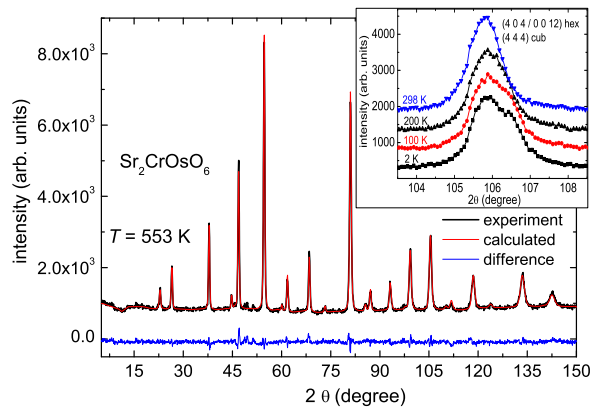}}
 \caption{(Color online) Rietveld refinements of the neutron powder data of
 Sr$_2$CrOsO$_6$ taken at 553\,K. The observed and calculated  diffraction
intensities as well as the difference pattern are shown. The upper
left inset shows a
  view along the [111]-direction revealing the rhombohedral structure of the double
 perovskite. The upper right inset shows the temperature dependence of the
 Sr$_2$CrOsO$_6$ (444) diffraction peak (cubic) splitting into a (404) and (0 0 12)
 peak
(hexagonal).}\label{Fig:neutron}
\end{figure}

Single phase polycrystalline samples of Sr$_2$CrOsO$_6$ have been
prepared from stoichiometric amounts of SrO$_{2}$, CrO$_2$ (both
Sigma-Aldrich, 99.999\%), and Os metal (Chempur 99.99\%) at
1273\,K in evacuated quartz tubes. The mixture was grinded
thoroughly, pressed into pellets and put into a corundum crucible.
The crucible was placed in a quartz tube which was further
evacuated and sealed under high vacuum. The quartz tube was heated
at 1273\,K for 2 days. The obtained product was further heated
under oxygen flow at 673\,K overnight to remove excess Os metal as
an impurity. Extensive x-ray and neutron diffraction analysis
confirms the high phase purity of the samples (see
Fig.~\ref{Fig:neutron}). The antisite disorder is negligible below
0.1\%.

\begin{table}[b]
\caption{\label{tab:moments} Measured and calculated (using the
generalized gradient approximation including spin-orbit coupling
(GGA+SO)) magnetic moments of Sr$_2$CrOsO$_6$ in
$\mu_{\textrm{B}}$/f.u. For a detailed discussion of the applied
band-structure calculation see e.g.
Ref.~\cite{Vaitheeswaran:05,Vaitheeswaran:06}. Note that the
neutron measurements are total moments per site. Calculated number
of $d$-holes at Os: $4.83$}
 \begin{ruledtabular}
  \begin{tabular}{l|ccc|ccc}
  method & Cr $m_{\textrm{S}}$ & & Cr $m_{\textrm{L}}$ & Os $m_{\textrm{S}}$  & & Os $m_{\textrm{L}}$ \\ \hline
  XMCD (300\,K)     &    &   &          & -0.17 & & 0.015  \\
  neutrons (298\,K) & & 1.92  &  & &  -0.12  &   \\
  neutrons (2\,K)  &  & 2.03  & & &  -0.73 &   \\
                      &     & &    &        &   &   \\
  GGA+SO (0\,K)       & 2.216 & & -0.024 & -1.214 & & 0.122    \\
  \end{tabular}
 \end{ruledtabular}
\end{table}

In Fig.~\ref{Fig:neutron} neutron powder diffraction data
collected with the instrument E9 (Ge monochromator selecting the
neutron wavelength $\lambda=1.7971$\,{\AA}) at the reactor BER II
of the Hahn-Meitner-Institut in Berlin is shown. The detailed
refinement results are presented in Table~\ref{tab:neutron}. From
data refinement one can say that the crystal structure is clearly
cubic above the magnetic transition temperature. At 550\,K the
crystal structure can still be refined by assuming a cubic
structure. At 300\,K the sample is already rhombohedral. The upper
right inset of Fig.~\ref{Fig:neutron} shows the temperature
evolution of the (444) diffraction peak in the cubic notation. At
low temperatures the splitting into two peaks ((404) and (0 0 12)
in the hexagonal notation) with an observed intensity ratio of
about 3 clearly evidences the rhombohedral structure. The
tolerance factor at low temperatures is still close to one
($\sim0.992$).

\begin{figure}[t]
\centering{%
\includegraphics[width=0.9\columnwidth,clip=]{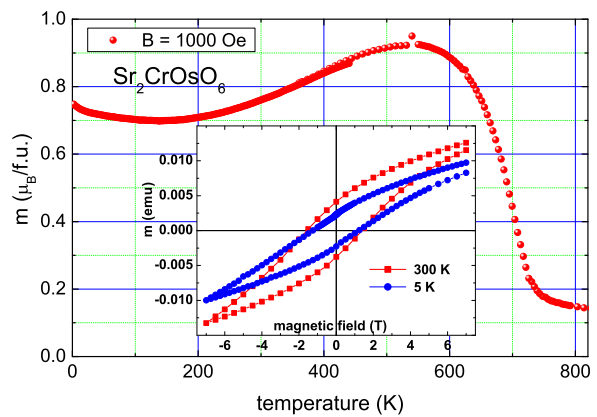}}
 \caption{(Color online) Magnetization vs temperature and hysteresis (at 5\,K and 300\,K) of
 Sr$_2$CrOsO$_6$ (measured in a superconducting quantum interference device (SQUID)).
 The sample has been cooled and field-oriented in
 diamagnetic LiNO$_3$ (melting point 480\,K) because of the high anisotropy field.}
 \label{Fig:SQUID}
\end{figure}

\begin{figure}[t]
\centering{%
\includegraphics[width=0.9\columnwidth,clip=]{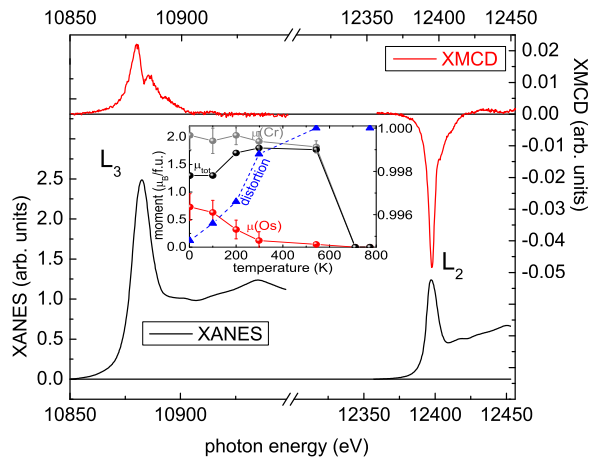}}
 \caption{(Color online) XANES (left axis) and XMCD signal (right axis)
 at the Os L$_2$ and L$_3$ edges in Sr$_2$CrOsO$_6$. The inset shows the
 element resolved magnetic moments of Cr ($\mu_{\text{Cr}}$),
 Os ($\mu_{\text{Os}}$), the total magnetic moment ($\mu_{\text{tot}}$)
 and the lattice distortion as determined from neutron scattering data
 refinement (see Table~\ref{tab:neutron}).}\label{Fig:xmcd}
\end{figure}

Figure~\ref{Fig:SQUID} shows the net magnetization and hysteresis
behavior of Sr$_2$CrOsO$_6$ as measured in a superconducting
quantum interference device (SQUID) magnetometer. $T_\textrm{C}$
is estimated to be about 725\,K. The magnetization maximum at
550\,K is a consequence of the different temperature evolution of
the Cr magnetic moments and the induced moments at the Os site
(see inset of Fig.~\ref{Fig:xmcd}). The drop of local magnetic
spin moment at the Os site with increasing temperature is clearly
not associated with a change of the insulating sample behavior. It
is most likely, that the unusual magnetic behavior is coupled to
the distortion of the lattice, as can be motivated from
Table~\ref{tab:neutron}. In the same temperature range where the
Os moment changes, the Os and Cr oxygen octahedra are markedly
twisted against each other (see topview of the crystal structure
in the inset of Fig.~\ref{Fig:neutron}, and the combined plot of
lattice distortion and magnetic moments vs.~temperature in the
inset of Fig.~\ref{Fig:SQUID}). Alternatively, one could invoke a
difficult to understand very strong ferromagnetic coupling between
the Cr ions along the chain Cr-O-Os-O-Cr. However, the finite
induced magnetic moment at the Os site, although reduced to
-0.05\,$\mu_{\text{B}}$ at 540\,K (for comparison: in
Sr$_2$FeReO$_6$ the total Re moment is only
-0.16\,$\mu_{\text{B}}$ at 5\,K \cite{DeTeresa:05}), survives up
to $T_\textrm{C}$ indicating the validity of the kinetic energy
gain model.


The XMCD measurements on the Os L$_{2,3}$ edges were performed at
the European Synchrotron Radiation Facility (ESRF) at beam line
ID12 \cite{Rogalev:01}.
As
shown in Fig.~\ref{Fig:xmcd} a very clear XMCD signal is observed,
signaling a strong local magnetic moment at the Os-site opposite
to the net magnetization. The experimentally determined value is
in good agreement with the neutron scattering result (see
table~\ref{tab:moments}).

\begin{figure}[b]
\centering{%
\includegraphics[width=0.9\columnwidth,clip=]{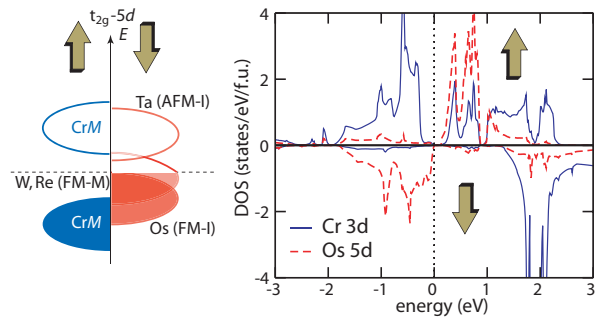}}
 \caption{(Color online) Left: Band structure sketch for different fillings of the 5$d$ spin-down
 band (as realized in the materials Sr$_2$Cr$M$O$_6$ with $M=$ Ta, W, Re,
 Os). Right:
 Band-structure calculation for Sr$_2$CrOsO$_6$ using the
generalized gradient approximation including spin-orbit coupling
(GGA+SO).}\label{Fig:bandstr}
\end{figure}

Fig.~\ref{Fig:bandstr} (right side) shows the site-resolved
density of states for the $d$-states obtained from band-structure
calculations, in the experimentally observed low-temperature
rhombohedral structure, using the generalized gradient
approximation (GGA) including spin-orbit coupling. The key result
is that Sr$_2$CrOsO$_6$ is, due to the completey filled Os
$t_{2\text{g}}$ spin down orbitals, indeed at the end of
spin-polarized metal-insulator transition (see sketch in
Fig.~\ref{Fig:bandstr}, left side). This is also in agreement with
conductivity measurements on the powder samples, showing about
10\,$\Omega$cm at room-temperature. Our combined experimental and
theoretical results show, that the kinetic-energy driven exchange
mechanism \cite{Sarma:00,Kanamori:01,Fang:01} is not only operable
in the itinerant range but in a wide range of conductivities.

In summary, we have unambiguously demonstrated, that an increase
of the number of $5d$ electrons leads to an increase of the
transition temperature in ferrimagnetic double perovskites
irrespective of the changes in conductivity. Fully spin-polarized
conductivity as indicated by the large induced magnetic moments
and supported further by band-structure calculations, can be
changed from half-metallic behavior (known in Sr$_2$CrWO$_6$ and
Sr$_2$CrReO$_6$) to insulating behavior in the high-Curie
temperature ferrimagnetic insulator Sr$_2$CrOsO$_6$.

The authors acknowledge the support by the DFG (Project Ns.~Al
560/4 and Ul 146/4), SSF and VR. SNIC is acknowledged for
providing computer facilities.

\begin{table*}[hb]
\caption{\label{tab:neutron} Results of the structure refinements
of Sr$_2$CrOsO$_6$. For the neutron powder data collected at 2\,K,
100\,K, 200\,K and 298\,K the crystal structure was refined in the
trigonal space group $R\overline{3}$. At the higher temperatures
540\,K and 770\,K the refinements were carried out in the cubic
space group $Fm\overline{3}m$. The residuals of the crystal and
magnetic structures are defined as $R_{N} =
\sum||F_{o}|-|F_{c}||/|F_{o}|$ and $R_{M}$ as $R_{M} =\sum
||I_{o}|- |I_{c}||/|I_{o}|$, respectively. For the refinements
several parameters were constrained to be equal. In these cases
the standard deviation is listed only for one of the equal
parameters. As distortion we define $c/(\sqrt{6}a)$, i.e.~the
ratio $c/a$ normalized to the value of a cubic structure in
hexagonal notation ($\sqrt{6}$). See also
\cite{Krockenberger:07}.}
\begin{ruledtabular}
\begin{tabular}{|l|l|l|l|l|l|l|}\hline
$T$\,[K] & 2 & 100 & 200 & 298 & 540 & 770  \\ \hline space group
& $R\overline{3}$ & $R\overline{3}$ & $R\overline{3}$ &
$R\overline{3}$ & $Fm\overline{3}m$ & $Fm\overline{3}m$ \\ \hline
$a$\,[{\AA}] & 5.5176(3) &  5.5170(4) & 5.5178(3) &  5.5181(6) &
7.8243(2) & 7.8455(3) \\ \hline

$c$\,[{\AA}] & 13.445(1) & 13.454(1) & 13.470(1) & 13.500(1) & --
& -- \\ \hline

$V$\,[{\AA}$^3$] & 354.47(4) &  354.64(4) &  355.16(4) & 355.98(8)
&  479.00(4) & 482.90(5) \\ \hline $c/a$ & 2.4367(4)
& 2.4386(3) & 2.4412(4) & 2.4464(8) & -- & -- \\
\hline distortion & 0.9948 & 0.9956 & 0.9966 & 0.9988 & 1 & 1 \\
\hline

$x_\text{O}$, $y_\text{O}$, $z_\text{O}$, site & 0.335(1), &
0.335(2), & 0.336(2), & 0.335(4), & 0.2511(5), 0, 0,  & 0.2511(5), 0, 0,  \\
& 0.1908(7), & 0.1900(7), & 0.1876(8), & 0.183(1), & 24$e$ & 24$e$\\
& 0.4169(6), 18$f$ & 0.4171(7), 18$f$ & 0.4171(8), 18$f$ & 0.417(2), 18$f$ & & \\
\hline

$x_\text{Sr}$, $y_\text{Sr}$, $z_\text{Sr}$, site & 0, 0 ,
0.251(1), 6$c$ &  0, 0 , 0.251(2), 6$c$ &  0, 0 , 0.251(2), 6$c$ &
0, 0 , 0.252(4), 6$c$ & $\frac{1}{4}$, $\frac{1}{4}$,
$\frac{1}{4}$, 8$c$  &  $\frac{1}{4}$, $\frac{1}{4}$,
$\frac{1}{4}$, 8$c$ \\ \hline

$x_\text{Os}$, $y_\text{Os}$, $z_\text{Os}$, site & 0, 0 ,
$\frac{1}{2}$, 3$b$ &  0, 0 , $\frac{1}{2}$, 3$b$ &  0, 0 ,
$\frac{1}{2}$, 3$b$ & 0, 0 , $\frac{1}{2}$, 3$b$ & 0, 0, 0,
4$a$  &  0, 0, 0, 4$a$ \\
\hline

$x_\text{Cr}$, $y_\text{Cr}$, $z_\text{Cr}$, site & 0, 0 , 0, 3$a$
&  0, 0 , 0, 3$a$ &  0, 0 , 0, 3$a$ & 0, 0 , 0, 3$a$ &
$\frac{1}{2}$, $\frac{1}{2}$, $\frac{1}{2}$,
4$b$  &  $\frac{1}{2}$, $\frac{1}{2}$, $\frac{1}{2}$, 4$b$ \\
\hline

$B$(Sr)\,[{\AA}$^2$] &
0.36(4) & 0.40(4) & 0.48(4) & 0.63(4) & 1.09(4) & 1.47(5) \\
\hline $B$(Cr,Os)\,[{\AA}$^2$] &
0.21(2)  & 0.22(2) &   0.25(2) &   0.30(3)  & 0.42(3) & 0.51(4) \\
\hline $B$(O)\,[{\AA}$^2$] & 0.45(3) & 0.47(3) & 0.50(3) & 0.67(3)
& 1.41(3) & 1.72(4) \\ \hline

$d$(Cr-O)\,[{\AA}] & 1.947(7)
& 1.948(8) &  1.946(9) & 1.95(2) &  1.948(4) & 1.953(4) \\
\hline

$d$(Os-O)\,[{\AA}] & 1.957(7) & 1.955(8) & 1.957(9) & 1.96(2) &
1.965(4) &  1.970(4) \\ \hline

$R_{\text{N}}$ & 0.039 & 0.040 &  0.040 & 0.037 &  0.034 &  0.039
\\ \hline\hline $\mu_{\text{Cr}}$\,[$\mu_{\text{B}}$] & 2.0(3) &  1.9(3) &
2.0(3) & 1.9(2) & 1.8(2) & 0 \\ \hline
$\mu_{\text{Os}}$\,[$\mu_{\text{B}}$] & -0.7(3) & -0.6(3) &
-0.3(2) & -0.1(2) & -0.05 & 0 \\ \hline
$R_{\text{M}}$ & 0.070 &  0.082 &  0.065 &  0.078 &  0.103 & -- \\
\hline
  \end{tabular}
  \end{ruledtabular}
\end{table*}

\end{document}